# Multifunctional photonic integrated circuit for diverse microwave signal generation, transmission and processing

Xihua Zou [1†*], Fang Zou [1†], Zizheng Cao [2†], Bing Lu [1], Xianglei Yan [1], Ge Yu [1], Xiong Deng [2], Bin Luo [1], Lianshan Yan [1], Wei Pan [1*], Jianping Yao [3*], and Antonius M. J. Koonen [2*]

[1] Center for Information Photonics and Communications, School of Information Science and Technology, Southwest Jiaotong University, Chengdu, 611756, China.
[2] Institute for Photonic Integration (the former COBRA Research Institute), Electro-Optics Communication Group, Eindhoven University of Technology, Eindhoven, 5600 MB, the Netherlands.
[3] Microwave Photonics Research Laboratory, School of Electrical Engineering and Computer Science, University of Ottawa, Ottawa, K1N 6N5, Canada.

†Authors contributed equally to this work.
*Corresponding authors. Emails: zouxihua@swjtu.edu.cn (X. Z.), wpan@swjtu.edu.cn (W. P.), jpyao@eecs.uottawa.ca (J. Y.) a.m.j.koonen@tue.nl (A. K.)

**Abstract**: Microwave photonics (MWP) studies the interaction between microwave and optical waves for the generation, transmission and processing of microwave signals (i.e., three key domains), taking advantages of broad bandwidth and low loss offered by modern photonics. Integrated MWP using photonic integrated circuits (PICs) can reach a compact, reliable and green implementation. Most PICs, however, are recently developed to perform one or more functions restricted inside a single domain. In this paper, as highly desired, a multifunctional PIC is proposed to cover the three key domains. The PIC is fabricated on InP platform by monolithically integrating four laser diodes and two modulators. Using the multifunctional PIC, seven fundamental functions across microwave signal generation, transmission and processing are demonstrated experimentally. Outdoor field trials for electromagnetic environment surveillance along an in-service high-speed railway are also performed. The success to such a PIC marks a key step forward for practical and massive MWP implementations.

# 1. Introduction

Microwave photonics (MWP), an emerging interdisciplinary field by bringing together microwave and optical waves [1-4], is capable of overcoming the current challenges and troubles in the generation, transmission and processing of microwave signals (i.e., three key domains of MWP), thanks to the inherently broadband and low-loss features of modern photonics. Traditionally, MWP systems using discrete components are fully exploited and implemented, for instance, in arbitrary waveform generator, wireless communications, signal processing and detection, and radar [5-9]. However, the scalability in both function and cost hinders the popularization of MWP. Hence photonic integrated circuit (PIC) leading to a compact and easy-to-handle implementation is needed to accelerate the integrated MWP's maturity [10-11], as it can greatly reduce the footprint and cost and considerably enhance the stability and energy efficiency.

Recently, a large number of PICs have been developed to implement different microwave devices or systems, including signal generators [12-16], filters [17-19], true-time delay lines and beamformers [20-23], signal processors [24-30], frontends and transceivers [31-35], and systems for signal characterization and interference cancellation [36-39]. Particularly, a few PICs have been developed to enable a generic, programmable processor or processor core [25-29] in the signal processing domain. However, these PICs are currently developed to perform one or more functions covering a single domain (i.e., microwave signal generation, transmission or processing). Namely, the PIC-assisted functions are restricted inside a single domain (see the overview listed in **Table 1**). The weakness behind the single-domain operation leads to long development time, high foundry cost, and thus limited applications. In addition, although PICs have been widely used in MWP field, they mostly have been only conceptually demonstrated in laboratory tests with well-conditioned laboratory environments or off-line conditions [25-30]. There exists a serious gap between laboratory demonstrations and practical applications, in which stability and robustness are two issues that must be addressed.

Therefore, we propose a multifunctional PIC capable of covering all the three key domains of MWP, the microwave signal generation, transmission and processing. The PIC is fabricated by monolithically integrating four tunable lasers, two modulators, and seven multimode interference (MMI) couplers. It is designed to enable a few multifunctional or switchable on-chip elements, modulation architectures, and optical signal-flows. Using the multifunctional PIC, seven fundamental microwave functions across the three domains have been enabled. The PIC is stable and robust for direct deployment in real-world applications, including electromagnetic environment surveillance for an in-service high-speed railway, broadband wireless communications, and remote high-definition (HD) video access. This work demonstrates a milestone step towards the evolution and maturity of the integrated MWP, and therefore toward the future massive applications.

**Table 1.** Overview of PICs for microwave signal generation, transmission and processing.

| Year & Ref. | Corresponding author & Affiliation | Three key domains of MWP (generation, transmission and processing) | | | Integration type | | Integrated element (number) | Service scenario | |
|---|---|---|---|---|---|---|---|---|---|
| | | Number | Domain | Function | Passive | Active | | Lab | Outdoor |
| 2010 [12] | A. M. Weiner Purdue U. | 1 | Generation | AWG | ☺ | | Micro-ring (8), MZI (8) | ☺ | ◉ |
| 2010 [20] | A. Meijerink U. Twente | 1 | Processing | Beamforming | ☺ | | ORR (4), splitters (3) | ☺ | ◉ |
| 2011 [17] | L. A. Coldren UCSB | 1 | Processing | Filter | ☺ | ☺ | SOA (21), PM (10), MMI (10) | ☺ | ◉ |
| 2012 [24] | F. X. Kartner MIT | 1 | Processing | ADC | ☺ | ☺ | MZM (1), ORR (2), PD (8), coupler (1) | ☺ | ◉ |
| 2013 [36] | J. S. Fandiño UPV | 1 | Processing | IFM | ☺ | | RAMZI (1) | ☺ | ◉ |
| 2014 [18] | S. B. Yoo UC Davis | 1 | Processing | Filter | ☺ | | PS (26), DC (18) | ☺ | ◉ |
| 2014 [21] | W. Shi McGill U. | 1 | Processing | TTD | ☺ | | WBG (1) | ☺ | ◉ |
| 2014 [34] | F. van Dijk III-V Lab | 1 | Generation | Transmitter | ☺ | ☺ | Laser (2), SOA (8), MMI (2), PD (2) | ☺ | ◉ |
| 2015 [25] | L. Zhuang Monash U. | 1 | Processing | Delay line, filter, Hilbert transform | ☺ | | Tunable M-Z couplers (array) | ☺ | ◉ |
| 2015 [37] | D. Marpaung Sydney U. | 1 | Processing | IFM | ☺ | | Silicon waveguide (1) | ☺ | ◉ |
| 2015 [13] | A. M. Weiner Purdue U. | 1 | Generation | AWG | ☺ | | ORR (16) | ☺ | ◉ |
| 2016 [38] | J. Azaña INRS-EMT | 1 | Processing | IFM | ☺ | | WBG (1) | ☺ | ◉ |
| 2016 [26] | J. Yao U. Ottawa | 1 | Processing | Differentiator, Hilbert transform, integrator | | ☺ | MMI (8), PM (12), SOA (9) | ☺ | ◉ |
| 2017 [29] | B. J. Eggleton U. Sydney | 1 | Processing | Filter | ☺ | | Chalcogenide waveguide (1) | ☺ | ◉ |
| 2017 [19] | J. S. Fandiño UPV | 1 | Processing | Filter | ☺ | ☺ | Laser (2), MZM (1), PM (1), Coupler (4), RAMZI (1), MZI (1), PD (4) | ☺ | ◉ |
| 2017 [22] | L. Zhou SJTU | 1 | Processing | TTD | ☺ | | ORR (array), MZI switch (array) | ☺ | ◉ |
| 2017 [30] | J. Yao U. Ottawa | 1 | Processing | Differentiator | ☺ | | MZI (1), MMI (1) | ☺ | ◉ |
| 2017 [14] | J. Hulme UCSB | 1 | Generation | Signal generator | ☺ | ☺ | Laser (2), PM (2), PD (1), DC (1) | ☺ | ◉ |
| 2017 [15] | K. Xu CUHK | 1 | Generation | Signal generator | ☺ | | ORR (2), MZI (1) | ☺ | ◉ |
| 2017 [27] | D. Pérez UPV | 1 | Processing | Multipurpose signal processor | ☺ | | 7 hexagonal MZI waveguide cells | ☺ | ◉ |
| 2018 [39] | M. P. Chang Princeton U. | 1 | Processing | Self-interference cancellation | ☺ | ☺ | Laser (2), SOA (3), PD (2), BPD (1) | ☺ | ◉ |
| 2018 [23] | D. J. Moss SUT | 1 | Processing | TTD | ☺ | | ORR (1) | ☺ | ◉ |
| 2018 [16] | M. Li CAS | 1 | Generation | OEO | ☺ | ☺ | Laser (1), Delay line (1), PD (1) | ☺ | ◉ |
| **2018, this paper** | | 3 | Generation Transmission Processing | Signal generator, analog links, filters, IFM, PRR measurement | ☺ | | Laser (4), PM (1), IM (1), MMI (7) | ☺ | ☺ (first) |

**Function.** ADC, analog to digital converter; AWG, arbitrary waveform generator; IFM, instantaneous frequency measurement; OEO, optoelectronic oscillator; PRR, pulse repetition rate; TTD, true time delay.

**Integrated element.** BPD, balanced photodetector; DC, directional coupler; IM, intensity modulator; MMI, multimode interferometer; MZI, Mach-Zehnder interferometer; MZM, Mach-Zehnder modulator; ORR, optical ring resonator; PD, photodetector; PM, phase modulator; PS, phase shifter; RAMZI, ring-assisted Mach-Zehnder interferometer; SOA, semiconductor optical amplifier; WBG, waveguide Bragg grating.

**Service scenario.** Lab, laboratory tests; Outdoor, outdoor real-world applications.

## 2. Multifunctional photonic integrated circuit (PIC)

### 2.1. PIC design and fabrication

The PIC is firstly designed and fabricated on the InP material platform offered by the European joint JePPIX platform (www.jeppix.eu). As shown in **Figure 1a**, it has four distributed Bragg reflection lasers (DBRLs) in parallel, one Mach-Zehnder intensity modulator (IM) and one phase modulator (PM). Each four-section DBRL (see **Figure 1b**) consists of two tunable distributed Bragg reflectors (TDBRs), one semiconductor optical amplifier (SOA) and one weak phase shift (PS). The SOA provides gain for laser emission by controlling the injection current, while the incorporation of the two TDBRs and the PS facilitates the wavelength tuning of each laser [40]. Seven 3-dB MMI couplers connect the DBRLs and modulators, to provide multifunctional architectures and thus diverse applications. In addition, DBRL 1 and DBRL 2 are in parallel connected with the IM via two complementary ports of an MMI coupler, while DBRL 4 is directly connected with the PM.

As depicted in **Figures 1c-1e**, the PIC has been delicately packaged with two optical ports (FC/APC) to obtain a switchable two-way transmission and four microwave GPPO ports to feed the IM and PM; it is then assembled with a printed circuit board (PCB) peripheral interface. The four DBRLs are tested with linewidths 3.3, 3.8, 4.3 and 2.5 MHz and with threshold currents of 13.2, 14.8, 15.1, and 16.1 mA. The IM and PM are characterized with an operation frequency up to 18 or 20 GHz (see **Section S1** of Supporting Information).

### 2.2. Multifunctional on-chip elements, architectures and signal-flows

The PIC is designed to be have multifunctional elements, modulation architectures, and switchable optical signal-flows. With the four-section body (see **Figure 1b**), each DBRL is playing a multifold role inside the PIC; it can operate as a single-wavelength laser for generating an optical carrier with tunable wavelength, an ultranarrow filter through optical injection locking, or a bandpass amplifier when biased slightly below the threshold current, as illustrated in **Figure 2**.

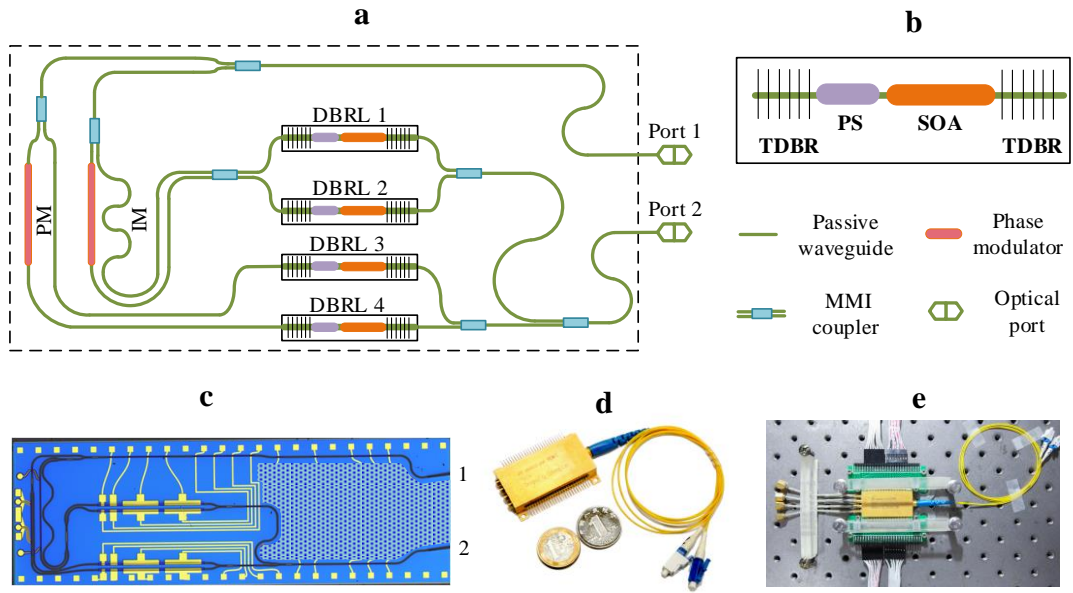

**Figure 1.** Layout and package of the multifunctional photonic integrated circuit. (a) Layout. (b) Four-section architecture of each DBRL. (c) Inside view of the packaged PIC. (d) External view of the packaged PIC. (e) Packaged PIC assembled with a printed circuit board (PCB) peripheral interface. (DBRL, distributed Bragg reflector laser; IM, intensity modulator; MMI: multimode interference; PM, phase modulator, PS, phase shift; SOA, semiconductor optical amplifier; TDBR, tunable distributed Bragg reflector)

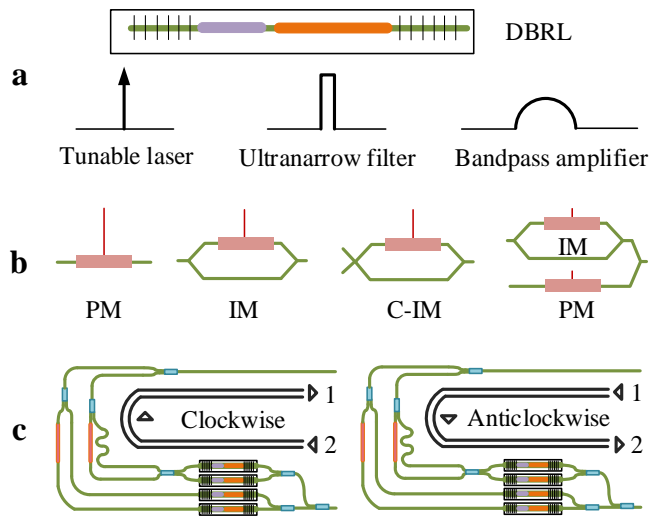

**Figure 2.** Multifunctional elements, architectures, and switchable signal-flows inside the photonic integrated circuit. (a) The on-chip TDBR-SOA-TDBR can be configured to act as a tunable laser, an ultranarrow filter, or a bandpass amplifier. (b) Four modulation architectures are available, including phase modulation, intensity modulation, complementary intensity modulation, and parallel phase and intensity modulation. (c) Both clockwise and anticlockwise optical signal-flows can be switched, providing a two-way transmission.

Four different modulation architectures are enabled in the PIC, as shown in **Figure 2b**. The phase modulation, intensity modulation, complementary intensity modulation, and parallel phase and intensity modulation are available, thanks to the flexible combination of the on-chip IM, PM, and MMI couplers.

Switchable optical signal-flows are also available in the PIC, since both optical links (without isolator or other nonreciprocal element) and microwave circuits allow a two-way transmission. Thus, the PIC can be configured to support either a clockwise optical signal-flow from Port 2 (input) to Port 1 (output) or an anticlockwise on from Port 1 (input) to Port 2 (output), as illustrated in **Figure 2c**. It is necessary to change the microwave input port of the on-chip IM or/and PM to retain high electrooptic modulation efficiency for the clockwise optical signal-flow or anticlockwise one. What we have to do is only to swap the microwave input port and the 50-$\Omega$ terminal port of the IM or/and PM. These multifunctional elements, architectures and switchable signal signal-flows enable the PIC to implement diverse applications across the three domains of microwave signal generation, transmission and processing.

## 3. Seven fundamental functions across three key domains

The multifunctional PIC is able to implement multiple fundamental functions across the three key domians of MWP (i.e., microwave signal generation, transmission and processing). In detail, seven microwave photonic functions are demonstrated here, including remote signal generator, analog intensity-modulation and phase-modulation MWP links, tunable bandpass and bandstop filters, microwave frequency measurement and pulse repetition rate measurement. The block diagrams and measured results are depicted in **Figure 3**, and more details can found in **Section S2** of Supporting Information.

*3.1. Signal generation domain: remote microwave signal generator*

To implement a remote microwave signal generator, the PIC is configured with DBRL 3, DBRL 4 and the PM, as shown in **Figure 3a**. An optical frequency comb (OFC) with a comb spacing of 11 GHz is coupled into the PIC. One comb line from the OFC is filtered out by DBRL 4 through optical injection locking and then modulated by a designed intermediate-frequency or baseband signal, while another comb line is selected by DBRL 3 in a similar way. The two comb lines are transmitted synchronously over a long single-mode fiber for remote distribution and generation. At the remote unit, beating the two comb lines generates microwave signals with user-defined waveform and tunable carrier frequency. For example, as illustrated on the right of **Figure 3b**,

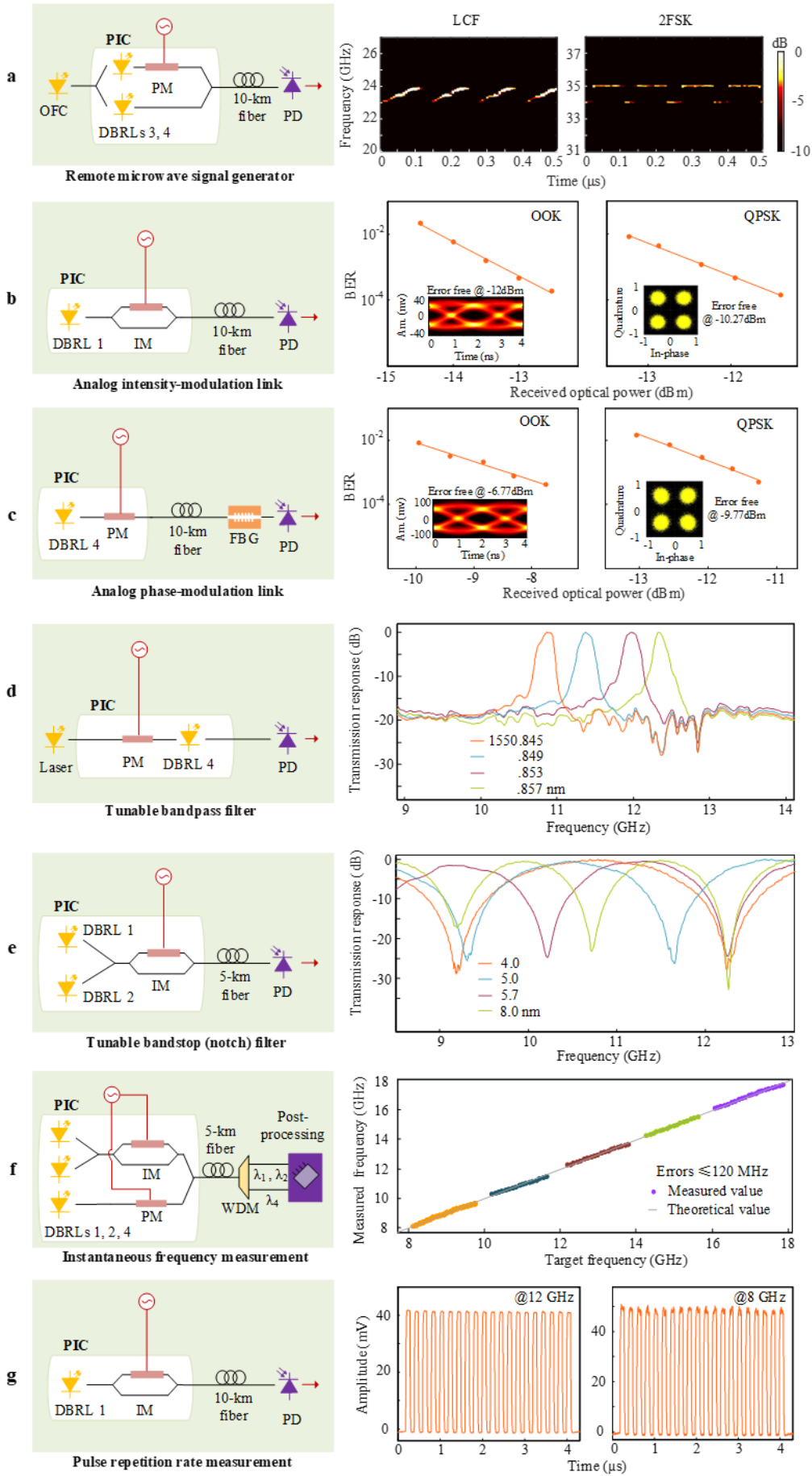

**Figure 3.** Seven fundamental microwave photonic functions enabled by the multifunctional PIC. (a) Remote microwave signal generator and the time-frequency distributions of the microwave signals generated with LCF and 2FSK. The LCF ranges from 23 to 24 GHz, while the 2FSK switches between 34 and 35 GHz. (b) Analog intensity-modulation link and (c) analog phase-modulation link; the BER curves, eye and constellation diagrams for the transmitted OOK and QPSK signals were presented. Signal processing: (d) bandpass filter and its tunable passbands from 10.8 to 12.4 GHz; (e) bandstop (notch) filter and its tunable stopbands (notches) from 9.3 to 12.3 GHz; (f) instantaneous frequency measurement system and measured microwave frequencies within the range of 8.1~17.9 GHz; (g) pulse repetition rate measurement system and measured pulse repetition rates for microwaves signals at 8 and 12 GHz.
(FBG, fiber Bragg grating; OFC, optical frequency comb; PD, photodetector; PIC, photonic integrated circuit; WDM, wavelength division multiplexer) one microwave signal with a linearly chirped frequency (LCF, from 23 to 24 GHz with a slope of 10 GHz µs$^{-1}$) and the other with binary frequency shift keying (2FSK, 34 or 35 GHz) are generated, when applying different intermediate-frequency signals to the on-chip PM. In addition, single-tone microwave signals with tunable carrier frequency at 22, 33, and 44 GHz have been generated (see **Figure S3**). It should be highlighted that the PIC ensures both short-term and long-term stability on the phase and amplitude of the generated microwave signals, which are extremely difficult to achieve in conventional two-branch architectures using discrete components (see discussions in **Section S3**, Supporting Information).

*3.2. Signal transmission domain: analog intensity-modulation and phase-modulation MWP links*

The PIC can be installed in analog intensity-modulation and phase-modulation MWP links, providing remote transmission of microwave signal. In the analog intensity-modulation MWP link, one laser (DBRL 1 or DBRL 2) is activated, and inside the on-chip IM the microwave signal to be transmitted is modulated on the optical carrier from the laser. For instance, an 8-GHz analog microwave signal carrying 500-Mbaud OOK (on-off keying) or QPSK (quadrature phase shift keying) baseband signal is applied, which is then transmitted over a 10-km single-mode fiber with the assistance of the PIC. In the central unit, the baseband signal is recovered via optical envelope detection and electronic down-conversion. The bit error rate (BER) curves and corresponding eye diagrams are recorded and shown in **Figure 3b**.

Likewise, in the phase-modulation link, the same microwave signal is applied to the PM inside the PIC, when DBRL 4 is switched on. As well known, linear phase demodulation is the critical step in the phase-modulation MWP link. Here, in the central unit the microwave signal is generated via phase-modulation to intensity-modulation conversion using an optical bandstop filter (actually a fiber Bragg grating with a 3-dB notch bandwidth of 0.08 nm) for phase demodulation. The BER curves and eye diagrams are recorded and shown in **Figure 3c**. It is clear that a BER of $10^{-4}$ is achieved at an optical power level around -10 dBm in both analog intensity-modulation and phase-modulation links when the microwave signal is carrying an OOK (500 Mbit s$^{-1}$) or QPSK (1 Gbit s$^{-1}$) signal.

## 3.3. Signal processing domain: tunable bandpass and bandstop filters, microwave frequency measurement and pulse repetition rate measurement

The PIC can also be employed for performing multiple functions in the microwave signal processing domain. In this section, tunable bandpass and bandstop filters, and systems for microwave frequency measurement and pulse repetition rate measurement will be demonstrated.

Both tunable bandpass and bandstop filters can be implemented by using the PIC. In the tunable bandpass filter, the PIC is configured to operate at an anticlockwise signal-flow mode and DBRL 4 is biased slightly below its threshold to act as a bandpass amplifier. The optical carrier of an external laser is phase modulated inside the on-chip PM, and then the bandpass amplifier provides a gain response to perform phase-modulation to intensity-modulation conversion. Tunable bandpass responses can be achieved only inside the gain response, when tuning the wavelength difference between the optical carrier and gain response. The measured passbands (see **Figure 3d**) indicate a tunable frequency range from 10.8 to 12.4 GHz, as the wavelength of the external laser is configured as 1550.845, 1550.849, 1550.853, and 1550.857 nm, respectively. Also, a tunable bandstop or notch filter can be configured inside the PIC as a two-tap transverse microwave filter, when introducing a 5-km fiber as a delay element. The center frequency of the tunable stopband or notch is determined by the wavelength spacing between DBRL 1 and DBRL 2. Thus, by tuning this wavelength spacing as 4.0, 5.0, 5.7, and 8.0 nm, a tunable range from 9.3 to 12.3 GHz has been observed for the stopband or notch (see **Figure 3e**).

Microwave frequency measurement [36-38] is of significance for civil and military applications, and the PIC enables us to implement such a measurement system. A dispersion-induced microwave power fading response from the PM branch of the PIC is incorporated with the periodic response of the two-tap microwave filter in the IM branch. As shown in **Figure 3f**, microwave frequency measurement has been realized within the range of 8.1~17.9 GHz without ambiguity, while frequency errors are estimated to be less than 120 MHz (see details in **Section S8** of Supporting Information).

For the measurement of the pulse repetition rate (PRR) or time-of-arrival (TOA) of a microwave signal [41], a system is designed by configuring the PIC with the IM and DBRL 1. The target microwave signal is remotely received by the PIC and transmitted to the central unit through a 10-km fiber. By incorporating a photodetector (PD) and a low-pass filter (or equivalently using a low-speed PD), the PRR can be measured in the central unit, totally independent of the carrier frequency. As shown in **Figure 3g**, the PRR at 10 MHz is successfully obtained from the target microwave signal without any prior information about the carrier frequency (8 or 12 GHz). Furthermore, the measurement to other PRRs (e.g., 50 and 100 MHz) of microwave signals at 8 and 12 GHz have also been demonstrated in **Figure S6**, while the carrier frequencies are unknown during the measurement.

More details about the seven functions referring to the three key domains of MWP can be found in **Sections S3-S9** of Supporting Information.

## 4. Real-world applications in the microwave signal transmission domain

Besides the experiments demonstrated in laboratory above, the PIC is fully packaged and assembled to support a host of real-world applications for microwave signal generation, transmission and processing in both indoor and outdoor scenarios for the first time. Concerning the microwave signal transmission using the PIC, in particular, the remote electromagnetic environment surveillance, broadband wireless communications, and remote HD video access have been demonstrated here.

### 4.1. Remote electromagnetic environment surveillance along an in-service high-speed railway

With the rapid growth of in-service high-speed railways all over the world (e.g., 26,869 km in-service high-speed lines in China and totally 41,908 km in the world by June 12, 2018 [42]), real-time electromagnetic environment surveillance is of great significance to ensure the safety and high operation efficiency of the high-speed railway network (see discussions in **Section S10** of Supporting Information). Using the PIC, a photonic system is designed to monitor the electromagnetic environment and interferences in the global system for mobile communications-railway (GSM-R, a radio system widely used to support high-speed train control & dispatching function). As shown in **Figures 4a-4c**, the multifunctional PIC is installed for implementing remote signal acquisition and transmission in either analog intensity-modulation or phase-modulation link, along the Chengdu-Chongqing High-speed Railway at a speed level of ~300 km h$^{-1}$ in Southwest China. At each test point, the remotely acquired radio signal is transmitted over a 10-km single-mode fiber to a central unit (e.g., railway station) where it is identified and analyzed both in the frequency domain and the temporal domain with high resolution. The spectra of the acquired radio signals and the cell identification (CI) of the frequency bands are then obtained, as illustrated in **Figures 4d** and **4e**. If no licensed CI is reco-

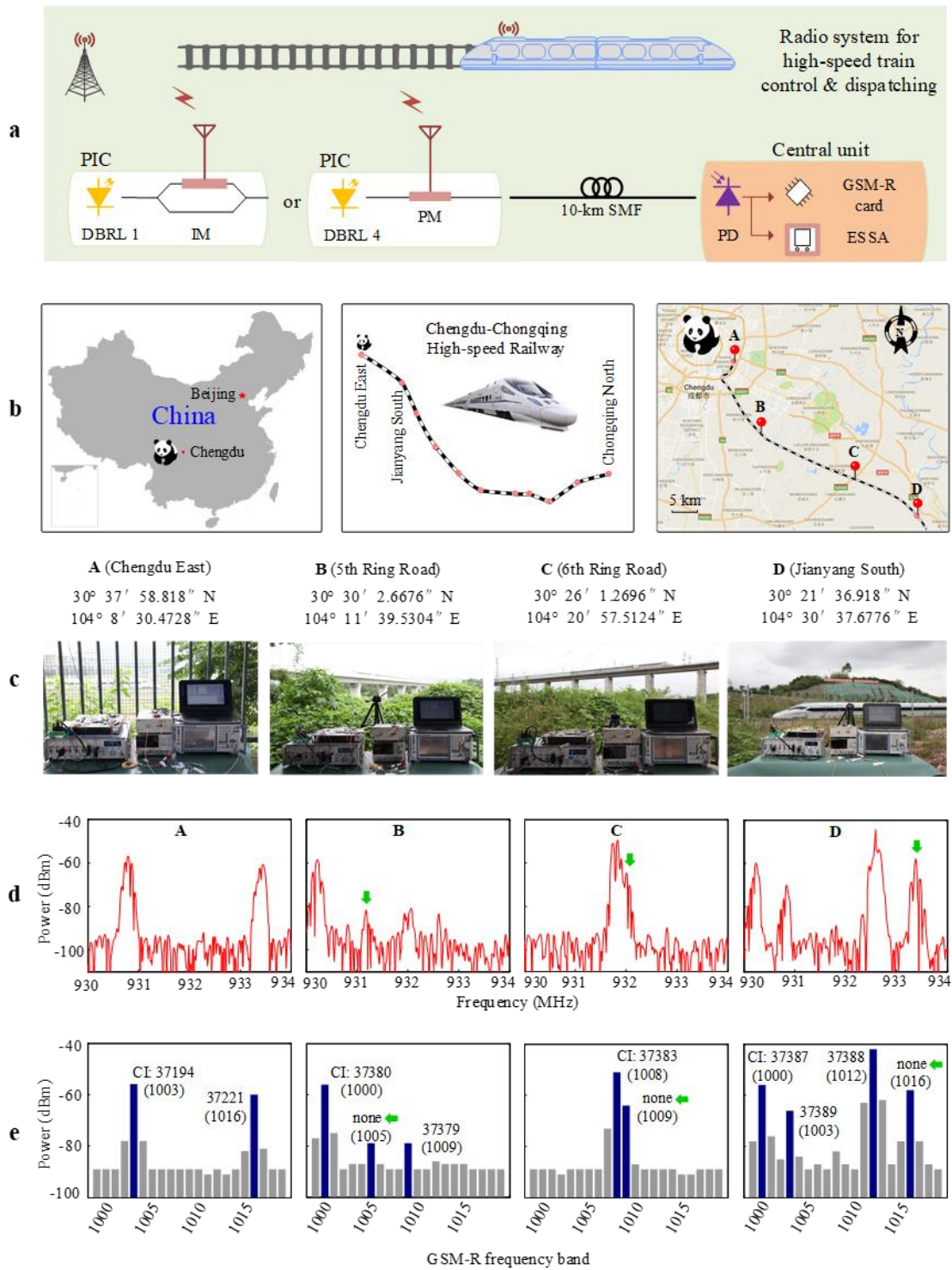

**Figure 4.** Field trials for electromagnetic environment surveillance along an in-service high-speed railway. (a) PIC-assisted photonic system for implementing remote signal acquisition and transmission. (b) Four test points distributed along the Chengdu-Chongqing High-speed Railway at a speed level of ~300 km h$^{-1}$ in Southwest China. (c) Locations and pictures of the field trials at the four test points. (d) Real-time radio spectra acquired from the four test points. (e) Cell identification (CI) and suspected electromagnetic interferences. The "green arrows" indicate suspected adjacent-channel or out-of-band interferences, as no licensed CI ('none') is recognized. The definitions of the CI (e.g., 37194) and frequency band (e.g., 1003) are described in **Section S10** of Supporting Information. (ESSA, electrical signal and spectrum analyzer; PIC, photonic integrated circuit)

-gnized (marked by "none" or "green arrows"), the radio components will be discriminated and alarmed as suspected adjacent-channel or out-of-band electromagnetic interferences (EMIs) from inter-modulation or foreign sources. Besides, the in-band or adjacent-channel interferences can also be effectively monitored from the eye and constellation diagrams, as the PIC is used for implementing remote microwave signal (GSM-R signal) acquisition and transmission. More results are shown in **Figure S8** of Supporting Information.

Consequently, based on the PIC-enabled function in the microwave signal transmission domain, a low-cost and real-time photonic system to electromagnetic environment surveillance has been established for promoting the safety and efficiency of the high-speed railway network. This photonic system can also be seamlessly compatible with next-generation wireless communication system, e.g., long-term evolution for railway (LTE-R). More details can be found in **Section S10** of Supporting Information and **Movie S1**.

*4.2. Broadband wireless communications and remote HD video access*

For the implementation of broadband wireless communications, the multifunctional PIC is embedded into a commercial time-division LTE (TD-LTE) system operating at 2.35 GHz, as shown on the top of **Figure 5**. It is installed in the remote radio unit (RRU) located on the roof of the college building, to transmit microwave signal in analog intensity-modulation link. Voice service between a cellular phone on the roof and a wired telephone connected to the base station, and video traffic between two cellular phones on the roof have been successfully demonstrated. (see **Section S11** of Supporting Information and **Movie S2**)

To provide remote HD video access, the multifunctional PIC is installed to transmit microwave signal in analog intensity-modulation link, as shown on the bottom of **Figure 5**. At first, real-time HD video signal (source) is converted into an HD-SDI (high-definition serial digital interface) signal at a data rate of 1.485 Gbit s$^{-1}$ and then upconverted to 9.1 GHz. The upconverted microwave signal carrying the HD video is applied to the on-chip IM to modulate the optical carrier from DBRL 1, and then remotely dispatched to user terminal over a 5-km fiber and 1.5-m free space. At the user terminal, the received microwave signal is down-converted and demodulated to recover the HD video signal, offering a remote high-fidelity video access. (see **Section S12** of Supporting Information and **Movie S3**).

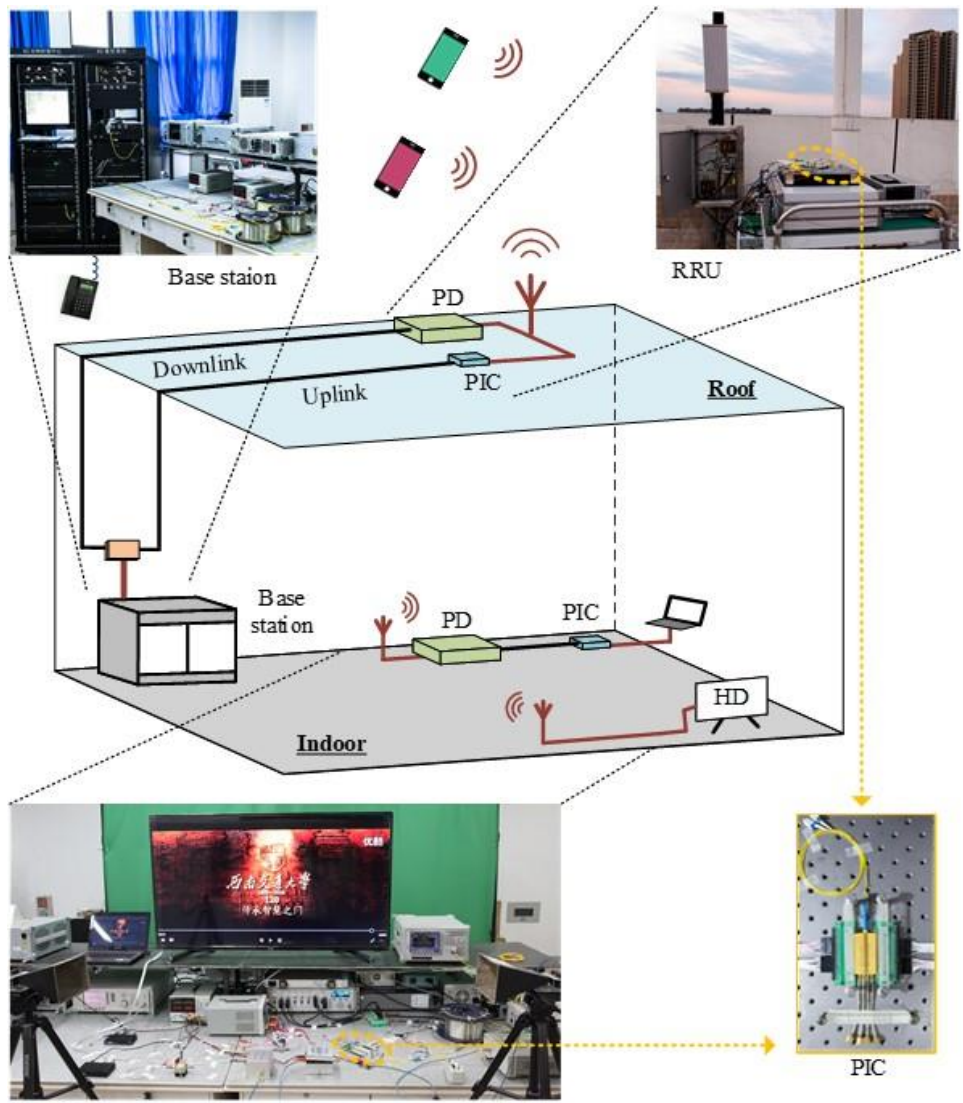

**Figure 5.** Field trials for broadband wireless communications (top) and remote HD video access (bottom), with the PIC-enabled microwave signal transmission. Videos about the two field trials can be found in **Movies S2** and **S3**. Inset: PIC installed in the two real-world application systems. (HD, high-definition video; PIC, photonic integrated circuit; RRU, remote radio unit)

## 5. Conclusion

We have proposed and demonstrated a multifunctional PIC for MWP. As listed in **Table 2**, the PIC has been designed to enable multiple fundamental microwave functions across the three key domains of MWP (i.e., microwave signal generation, transmission and processing), not a single function in previous reports. They are compelling for a wide range of applications, including the remote microwave signal generator, analog intensity-modulation and phase-modulation MWP links, tunable microwave bandpass and bandstop filters, microwave frequency measurement and pulse repetition rate measurement.

Then the PIC has been directly deployed for remote signal transmission in real-world MWP applications, such as electromagnetic environment surveillance for an in-service high-speed railway, broadband wireless communications, and remote HD video access. To the best of our knowledge (see the overview in **Table 1**), this work can be considered as the first demonstration to a multifunctional PIC for diverse applications across the microwave signal generation, transmission and processing, as well as real-world applications integrated MWP in both indoor and outdoor scenarios.

**Table 2.** Summary of the multifunctional PIC for the seven MWP functions.

| MWP domain | Available function | Multifunctional elements and configurations | Laboratory tests | Real-world applications |
|---|---|---|---|---|
| Generation | Remote microwave signal generator | Ultranarrow filter (DBRL); Phase modulation; Clockwise signal-flow. | ☺ | |
| Transmission | Analog intensity-modulation link | Tunable laser (DBRL); Intensity modulation; Clockwise signal-flow. | ☺ | ☺ a), b), c) |
| | Analog phase-modulation link | Tunable laser (DBRL); Phase modulation; Clockwise signal-flow. | ☺ | ☺ a) |
| Processing | Tunable bandpass filter | Bandpass amplifier (DBRL); Phase modulation; Anticlockwise signal-flow. | ☺ | |
| | Tunable bandstop filter | Tunable laser (DBRL); Intensity modulation; Clockwise signal-flow. | ☺ | |
| | Microwave frequency measurement system | Tunable laser (DBRL); Parallel modulation; Clockwise signal-flow. | ☺ | |
| | Pulse repetition rate measurement system | Tunable laser (DBRL); Intensity modulation; Clockwise signal-flow. | ☺ | |

**a)**. High-speed railways; **b)**. broadband wireless communications; **c)**. remote HD video access.

**Supporting Information**

Supporting Information is available from the Wiley Online Library or from the authors.

**Section S1.** Characterization of the multifunctional photonic integrated circuit.

**Section S2.** Multifunctional elements, architectures, and signal-flows for microwave signal generation, transmission and processing.

**Section S3.** Remote microwave signal generator.

**Section S4.** Analog intensity-modulation microwave photonic link.

**Section S5.** Analog phase-modulation microwave photonic link.

**Section S6.** Tunable microwave photonic bandpass filter.

**Section S7.** Tunable microwave photonic bandstop filter.

**Section S8.** Photonic microwave frequency measurement.

**Section S9.** Photonic pulse repetition rate measurement.

**Section S10.** Electromagnetic environment surveillance for in-service high-speed railways.

**Section S11.** Broadband wireless communications.

**Section S12.** Remote high-definition video access.

**Section S13.** Performance and specification comparison with other PICs.

**Movie S1**. Field trial: electromagnetic environment surveillance for in-service high-speed railways.

**Movie S2**. Field trial: broadband wireless communications.

**Movie S3**. Field trial: remote high-definition video access.

**Acknowledgments**

The work was supported in part by the National High Technology Research and Development Program of China (2015AA016903), the National Natural Science Foundation of China (61775185, 61735015), and the Sichuan Science and Technology Program (2018HH0002). X. Zou was supported by the Research Fellowship of the Alexander von Humboldt Foundation, Germany.

**Conflict of Interest**

The authors declare no conflict of interest.

**Keywords**

integrated microwave photonics, photonic integrated circuit, microwave signal generation/transmission/processing, electromagnetic environment surveillance, broadband wireless communications.

## References

[1]. A. J. Seeds and K. J. Williams, *J. Lightwave Technol.* **2006**, *24*, 4628.

[2]. R. Minasian, *IEEE Trans. Microw. Theory Tech.* **2006**, *54*, 832.

[3]. J. Capmany and D. Novak, *Nat. Photon.* **2007**, *1*, 319.

[4]. T. Berceli and P. R. Herczfeld, *IEEE Trans. Microw. Theory Tech.* **2010**, *58*, 2992.

[5]. Z. Jia, J. Yu, G. Ellinas, and G. K. Chang, *J. Lightwave Technol.* **2007**, *25*, 3452.

[6]. S. Koenig, D. Lopez-Diaz, J. Antes, F. Boes, R. Henneberger, A. Leuther, A. Tessmann, R. Schmogrow, D. Hillerkuss, R. Palmer, and T. Zwick, *Nat. Photon.* **2013**, *7*, 977.

[7]. P. Ghelfi, F. Laghezza, F. Scotti, G. Serafino, A. Capria, S. Pinna, D. Onori, C. Porzi, M. Scaffardi, A. Malacarne, and V. Vercesi, *Nature* **2014**, *507*, 341.

[8]. T. Nagatsuma, G. Ducournau, and C. C. Renaud, *Nat. Photon.* **2016**, *10*, 371.

[9]. X. Zou, B. Lu, W. Pan, L. Yan, A. Stöhr, and J. Yao, *Laser Photon. Rev.* **2016**, *10*, 711.

[10]. D. Marpaung, C. Roeloffzen, R. Heideman, A. Leinse, S. Sales, and J. Capmany, *Laser Photon. Rev.* **2013**, *7*, 506.

[11]. L. R. Chen, *J. Lightwave Technol.* **2017**, *35*, 824.

[12]. M. H. Khan, H. Shen, Y. Xuan, L. Zhao, S. Xiao, D. E. Leaird, A. M. Weiner, and M. Qi, *Nat. Photon.* **2010**, *4*, 117.

[13]. J. Wang, H. Shen, L. Fan, R. Wu, B. Niu, L. T. Varghese, Y. Xuan, D. E. Leaird, X. Wang, F. Gan, and A. M. Weiner, *Nat. Commun.* **2015**, *6*, 5957.

[14]. J. Hulme, M. J. Kennedy, R. L. Chao, L. Liang, T. Komljenovic, J. W. Shi, B. Szafraniec, D. Baney, and J. E. Bowers, *Opt. Express* **2017**, *25*, 2422.

[15]. X. Wu, K. Xu, W. Zhou, C. W. Chow, and H. K. Tsang, *IEEE Photon. Technol. Lett.* **2017**, *29*, 1896.

[16]. J. Tang, T. Hao, W. Li, D. Domenech, R. Baños, P. Muñoz, N. Zhu, J. Capmany, and M. Li, *Opt. Express* **2018**, *26*, 12257.

[17]. E. J. Norberg, R. S. Guzzon, J. S. Parker, L. A. Johansson, and L. A. Coldren, *J. Lightwave Technol.* **2011**, *29*, 1611.

[18]. B. Guan, S. S. Djordjevic, N. K. Fontaine, L. Zhou, S. Ibrahim, R. P. Scott, D. J. Geisler, Z. Ding, and S. B. Yoo, *IEEE J. Sel. Top. Quantum Electron.* **2014**, *20*, 359.

[19]. J. S. Fandiño, P. Muñoz, D. Doménech, and J. Capmany, *Nat. Photon.* **2017**, *11*, 124.

[20]. A. Meijerink, C. G. Roeloffzen, R. Meijerink, L. Zhuang, D. A. Marpaung, M. J. Bentum, M. Burla, J. Verpoorte, P. Jorna, A. Hulzinga, and W. van Etten, *J. Lightwave Technol.* **2010**, *28*, 3.

[21]. W. Shi, V. Veerasubramanian, D. Patel, and D. Plant, *Opt. Lett.* **2014**, *39*, 701.

[22]. X. Wang, L. Zhou, R. Li, J. Xie, L. Lu, K. Wu, and J. Chen, *Optica* **2017**, *4*, 506.

[23]. X. Xu, J. Wu, T. G. Nguyen, T. Moein, S. T. Chu, B. E. Little, R. Morandotti, A. Mitchell, and D. J. Moss, *Photon. Res.* **2018**, *6*, B30.

[24]. A. Khilo, S. J. Spector, M. E. Grein, A. H. Nejadmalayeri, C. W. Holzwarth, M. Y. Sander, M. S. Dahlem, M. Y. Peng, M. W. Geis, N. A. DiLello, J. U. Yoon, A. Motamedi, J. S. Orcutt, J. P. Wang, C. M. Sorace-Agaskar, M. A. Popovic, J. Sun, G. Zhou, H. Byun, J. Chen, J. L. Hoyt, H. I. Smith, R. J. Ram, M. Perrott, T. M. Lyszczarz, E. P. Ippen, and F. X. Kartner, *Opt. Express* **2012**, *20*, 4454.

[25]. L. Zhuang, C. G. Roeloffzen, M. Hoekman, K. J. Boller, and A. J. Lowery, *Optica* **2015**, *2*, 854.

[26]. W. Liu, M. Li, R. S. Guzzon, E. J. Norberg, J. S. Parker, M. Lu, L. A. Coldren, and J. Yao, *Nat. Photon.* **2016**, *10*, 190.

[27]. D. Pérez, I. Gasulla, L. Crudgington, D. J. Thomson, A. Z. Khokhar, K. Li, W. Cao, G. Z. Mashanovich, and J. Capmany, *Nat. Commun.* **2017**, *8*, 636.

[28]. D. A. B. Miller, *Photon. Res.* **2013**, *1*, 1.

[29]. A. Choudhary, B. Morrison, I. Aryanfar, S. Shahnia, M. Pagani, Y. Liu, K. Vu, S. Madden, D. Marpaung, and B. J. Eggleton, *J. Lightwave Technol.* **2017**, *35*, 846.

[30]. W. Liu, W. Zhang, and J. Yao, *J. Lightwave Technol.* **2017**, *35*, 2487.

[31]. R. C. Hsu, A. Ayazi, B. Houshmand, and B. Jalali, *Nat. Photon.* **2007**, *1*, 535.

[32]. V. S. Ilchenko, A. B. Matsko, I. Solomatine, A. A. Savchenkov, D. Seidel, and L. Maleki, *IEEE Photon. Technol. Lett*. **2008**, *20*, 1600.

[33]. T. R. Clark and R. Waterhouse, *IEEE Microw. Mag.* **2011**, *12*, 87.

[34]. F. van Dijk, G. Kervella, M. Lamponi, M. Chtioui, F. Lelarge, E. Vinet, Y. Robert, M. J. Fice, C. C. Renaud, A. Jimenez, and G. Carpintero, *IEEE Photon. Technol. Lett*. **2014**, *26*, 965.

[35]. X. Zou, W. Bai, W. Chen, P. Li, B. Lu, G. Yu, W. Pan, B. Luo, L. Yan, and L. Shao, *J. Lightwave Technol.* **2018**, *36*, 4337.

[36]. J. S. Fandiño and P. Muñoz, *Opt. Lett*. **2013**, *38*, 4316.

[37]. M. Pagani, B. Morrison, Y. Zhang, A. Casas-Bedoya, T. Aalto, M. Harjanne, M. Kapulainen, B. J. Eggleton, and D. Marpaung, *Optica* **2015**, *2*, 751.

[38]. M. Burla, X. Wang, M. Li, L. Chrostowski, and J. Azaña, *Nat. Commun*. **2016**, *7*, 13004.

[39]. M. P. Chang, E. C. Blow, M. Z. Lu, J. J. Sun, and P. R. Prucnal, *IEEE Trans. Microw. Theory Tech*. **2018**, *66*, 596.

[40]. A. J. Ward, D. J. Robbins, G. Busico, E. Barton, L. Ponnampalam, J. P. Duck, N. D. Whitbread, P. J. Williams, D. C. Reid, A. C. Carter, and M. J. Wale, *IEEE J. Sel. Top. Quantum Electron*. **2005**, *11*, 149.

[41]. J. B. Tsui, *Microwave Receivers with Electronic Warfare Applications* (Wiley, New York, NY, USA, **1986**), chapter 3.

[42]. International Union of Railways (UIC), *High speed lines in the world (summary)*. Available at https://uic.org/IMG/pdf/20180612a_high_speed_lines_in_the_world.pdf (June 12, **2018**). Accessed: August 30, 2018.